\documentclass[aps,prb,showpacs,floatfix,preprint]{revtex4-1}
\usepackage{amsmath} 
\usepackage{chemarr} 
\usepackage{amssymb} 
\usepackage{graphicx}
\usepackage{color}

\begin{document}

\title{Network analysis of the performance of organic photovoltaic cells: The open circuit voltage and the zero current efficiency}

\author{Mario Einax}
\email{meinax@al.tau.ac.il} 
\affiliation{School of Chemistry, Tel Aviv University, Tel
Aviv 69978, Israel}
\author{Abraham Nitzan}
\email{nitzan@post.tau.ac.il}
\affiliation{School of Chemistry, Tel Aviv University, Tel
Aviv 69978, Israel}

\date{\today}

\begin{abstract}
Photovoltaic energy conversion in photovoltaic cells has been
analyzed by the detailed balance approach or by thermodynamic
arguments. Here we introduce a network representation to analyze the
performance of such systems once a suitable kinetic model
(represented by a master equation in the space of the different
system states) has been constructed. Such network representation
allows one to decompose the steady state dynamics into cycles,
characterized by their cycle affinities. The maximum achievable
efficiency of the device is obtained in the zero affinity limit.
This method is applied to analyze a microscopic model for a bulk
heterojunction organic solar cell that includes the essential
optical and interfacial electronic processes that characterize this
system, leading to an explicit expression for the theoretical
efficiency limit in such system. In particular, the deviation from
Carnot's efficiency associated with the exciton binding energy is
quantified.
\end{abstract}


\maketitle

\section{Introduction}
\label{sec:intro}
The quest to improve the efficiency of solar energy conversion is the
focus of intensive current research. \cite{Nelson:2003,Wuerfel:2009,Nayak/etal:2012}
In particular, considerable attention has focused recently on organic solar cells,
where advantageous low manufacturing cost is still counterbalanced by
a relatively low energy conversion yield, associated with the fact
that light absorption in such low dielectric permittivity materials
forms excitons, that is electron-hole pairs,\cite{Bruetting:2012} that
require extra energy for dissociation.\cite{Denner/etal:2009,Chen/etal:2009,Bredas/etal:2009,Deibel/Dyakonov:2010,Nicholson/Castro:2010,Thompson/etal:2011,Nelson:2011,Camaioni/Po:2013,Seki/etal:2013}

Such energy conversion studies naturally involve question concerning
efficiency,\cite{Nelson:2011,Potscavage/etal:2009,Wagenpfahl/etal:2010,Koster/etal:2012,Gruber/etal:2012}
in particular the possible existence of
fundamental limits on this efficiency.\cite{Shockley/Queisser:1961,Henry:1980,Landsberg/Tonge:1980,Giebink/etal:2011,Schaber/Sariciftci:2013,Green:2012}
Obviously, the efficiency of any individual photovoltaic system intimately depends on its
structure, but much as is done for heat engines, it is of interest to
understand it on the generic level which starts with the determination
of the maximum efficiency and follows by identifying and analyzing
processes that reduce it. The seminal work of Shockley and Queisser
(SQ)\cite{Shockley/Queisser:1961} is a prominent example. In that work, a thermodynamic analysis
of semiconductor (SC)-based solar cells is carried out under the
assumptions that \emph{(a)} all photons with energies larger than the SC band
gap are absorbed, and \emph{(b)} the only source of loss is the radiative
recombination of e-h pairs (an unavoidable process whose existence
follows from the principle of detailed balance). With these model
assumptions, and using thermodynamic considerations formulated in
terms of the detailed balance principle, SQ has provided a simple
analysis of the maximal ensuing cell efficiency. Several works, see for example Refs.~\onlinecite{Sylvester-Hivid/etal:2004,Rau:2007,Kirchartz/Rau:2008,Kirchartz/etal:2009a,Vandewal/etal:2009},
have extended the SQ analysis to more complex models, e.~g.,
organic photovoltaic (OPV) cells.\cite{Seki/etal:2013,Koster/etal:2012,Giebink/etal:2011,Schaber/Sariciftci:2013,Sylvester-Hivid/etal:2004,Kirchartz/etal:2009a,Vandewal/etal:2009,Miyadera/etal:2014}
Others have formulated abstractions of the SQ model (sometimes with
generalizations that account for carrier non-radiative recombination)
in order to study its kinetics and thermodynamics foundation.\cite{Nelson/etal:2004,Markvart:2008,Rutten/etal:2009,Einax/etal:2011,Einax/etal:2013,Wang/Wu:2012}
Recent works have also studied the possible implications of quantum
coherence in the quantum analogues of such kinetic models.\cite{Scully:2010,Kirk:2011,Scully:2011,Goswami/Harbola:2013}

At the core of many of these generic approaches is the use of
thermodynamics to analyze energy exchange and conversion processes in
the limit of vanishing rates. Such analysis can provide generic
results for maximal efficiencies at the cost of being limited to zero
power processes. Consideration of such systems under finite power
operation requires more detailed information about the underlying rate
processes. This has been done for specific model systems, see e.~g. Ref.~\onlinecite{Rutten/etal:2009},
however it is of interest to find a general formulation and generic
principles that underline the analysis of such situations. Obviously,
such an analysis should reduce to its thermodynamic counterpart in the
limit of zero rates (that is, equilibrium) and power.

In this paper we formulate this task in the framework of network
theory as applied to steady state systems.\cite{Hill:1966,Schnakenberg:1976,Zia/Schmittmann:2007,Andrieux/Gaspard:2007,Gaspard:2010,Altaner/etal:2012,Seifert:2012}
Inspired by the Kirchoff laws,\cite{Kirchhoff:1847} applications of this theory to the performance
analysis of chemical reaction networks are well known in diverse areas
such as chemical engineering\cite{Andrieux/Gaspard:2004} and chemical biology,\cite{Gerritsma/Gaspard:2010} but we are not
aware of such work on photovoltaic systems. We will limit ourselves to
the open circuit (OC), reversible operation limit, leaving dynamic
considerations to a subsequent publication. When applied (Section~\ref{sec:2-level})
to the simplest $2$-level model of Refs.~\onlinecite{Nelson/etal:2004} and \onlinecite{Rutten/etal:2009} this framework yields
a formalism similar to that considered in these papers. The strength
of this approach becomes apparent in more complex models as we show in
the subsequent consideration (Section~\ref{sec:BHJ-OPV_model}) of the thermodynamic
efficiency limit in the simplest ($6$-level) kinetic model\cite{Einax/etal:2011,Einax/etal:2013} for an
organic bulk heterojunction (BHJ) solar cell. (While we consider this
model in detail, it is made evident that this description can be
applied in far more complex situations.) The system dynamics is
described by a kinetics scheme derived using a lattice gas
approach,\cite{Einax/etal:2010a,Einax/etal:2010b,Dierl/etal:2011,Dierl/etal:2012}
similar in spirit to previous work\cite{Wagenpfahl/etal:2010,Sylvester-Hivid/etal:2004,Burlakov/etal:2005,Ruehle/etal:2011}
that use a master equation approach to analyze cell dynamics. In the graph
theory approach this kinetic scheme is represented by a graph that
comprises nodes (corresponding to states) and edges (representing
transitions between states), on which fluxes associated with the
non-equilibrium dynamics flow along interconnected linear and cyclical
paths. In this scheme, the observed macroscopic currents (average
currents of macroscopic variables) through the systems, are linked
through their circular counterparts to the microscopic transitions
between individual states. It has been shown by Schnakenberg\cite{Schnakenberg:1976} that
for each cycle an associated entropy production (called affinity of
cycle) can be obtained as the ratio between the product of all
transition rates in the forward direction and the corresponding
product of transition rates in the reversed direction. Then, the upper
efficiency limit of a large class of systems follows straightforwardly
by setting the cycle affinity of a basic cycle (that contains the
photovoltaic operation of the device) to zero. Specifying to BHJ-OPV
cells, this analysis shows that when exciton binding energies are
non-negligible the molecular heat engines operates with an efficiency
which is fundamentally lower than the Carnot efficiency. This finding
recovers the numerical observation in Ref.~\onlinecite{Einax/etal:2011} and is compatible with
the result obtained from the second law of thermodynamics in Ref.~\onlinecite{Giebink/etal:2011}.
As expected, in the limit of zero exciton binding the theoretical
limit approaches the universal upper bound given by the Carnot
efficiency.

\section{The 2-level photovoltaic model}
\label{sec:2-level}
\begin{figure}[h!]
\centering
\includegraphics[width=0.95\textwidth]{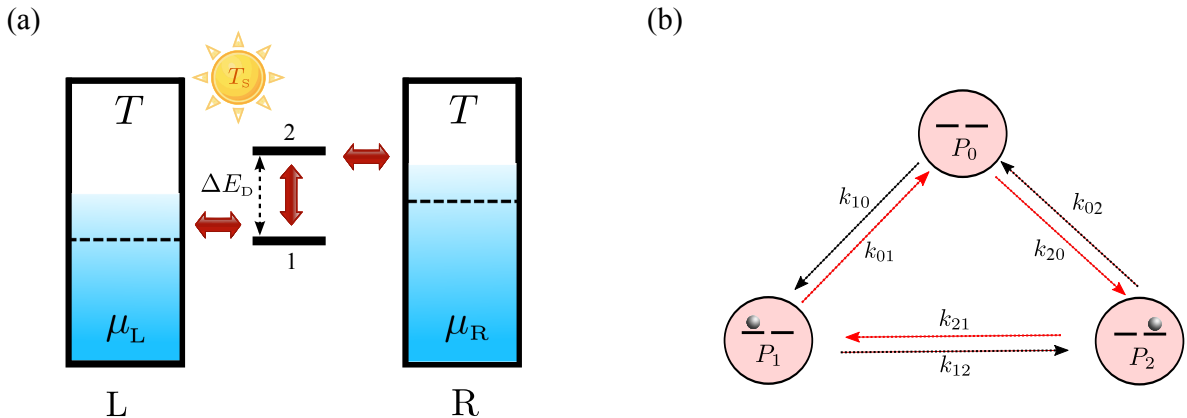}
\caption{A spinless two-levels, $3$-state model of a solar device that comprises two
  metal electrodes and a two-level molecule. Levels $1$ and $2$ are
  coupled to the left and right electrodes, respectively. The molecule
  can be in states $0=|0,0\rangle$, $1=|1,0\rangle$ and
  $2=|0,1\rangle$ where $|n_1, n_2\rangle$ is a state with $n_1$
  electrons in level $1$ and $n_2$ electrons in level $2$ (double
  occupancy is not allowed).}
\label{fig:fig1}
\end{figure}

As in Refs.~\onlinecite{Nelson/etal:2004,Rutten/etal:2009,Baruch:1985},
we consider a photovoltaic device comprising a
two level system situated between two external contacts, $L$ and $R$
[see Fig.\ref{fig:fig1}(a)], so that level $1$ is coupled only to the
left electrode while level $2$ sees only the right electrode.  For
simplicity we disregard the electron spin and exclude double occupancy
of the $2$-level system. This device can thus be in three states:
$0$-vacant, $1$-electron in level $1$ and $2$-electron in level $2$,
that constitute a simple cyclical network [Fig.\ref{fig:fig1}(b)] in
which each vortex represent a state and each edge connecting two
vortices corresponds to a pair of forward and back rates
\begin{align}
\label{eq:fb_path}
0 & \xrightleftharpoons[k_{01}]{k_{10}} 1 \xrightleftharpoons[k_{12}]{k_{21}} 2 \xrightleftharpoons[k_{20}]{k_{02}} 0
\end{align}
Under conditions that lead to equilibrium at long time, the ratios between these rates are
determined by the ambient temperature $T$ , the level energies $E_1$, $E_2$ and the chemical
potential $\mu$ that characterizes electrons in the metal electrodes, and are given by the
detailed balance relations
\begin{align}
\label{eq:DBC_2-level_eq}
\frac{k_{10}}{k_{01}} &= e^{-\beta (E_1-\mu)}\, ; \quad \frac{k_{21}}{k_{12}} = e^{-\beta (E_2-E_1)}\, ; \quad
\frac{k_{20}}{k_{02}} = e^{-\beta (E_2-\mu)}\, ,
\end{align}
where $\beta=1/k_{\rm\scriptscriptstyle B} T$ is the inverse thermal energy and
$k_{\rm\scriptscriptstyle B}$ is the Boltzmann constant. Note that the rates $k_{21}$ and
$k_{12}$ can originate from radiative transition (thermal radiation)
as well as non-radiative processes, both characterized by the ambient
temperature $T$. At equilibrium all fluxes vanish, $J_{ji} =k_{ji}
P_i^{\rm eq} - k_{ij} P_j^{\rm eq}$, where $P_j$ is the probability
that the system is in state $j$. A cyclical network of this property
is characterized by the identity
\begin{align}
\label{eq:path_ratio_2-level}
\frac{k_{02} k_{21} k_{10}}{k_{20} k_{12} k_{01}} = 1
\end{align}
that is satisfied by the ratio between forward and backward rates in a reaction loop,
provided that these rates sustain a state of zero loop current.

In an operating photovoltaic cell the system is taken out of this
equilibrium in two ways: \emph{(a)} Radiative pumping (an damping) is
affected on the $1$-$2$ transition. In standard models of photovoltaic
cells this pumping is represented by an effective temperature
$T_{\rm\scriptscriptstyle S} =1/k_{\rm\scriptscriptstyle B} \beta_{\rm\scriptscriptstyle S}$ (``sun temperature''\cite{note:EN_1}).
With the coupling scheme (\ref{eq:fb_path}) this leads to electron current from the left
to the right electrode, however this short circuit current does not
perform any useful work unless \emph{(b)} an opposing voltage bias $V=\Delta
\mu /e$ is set between the two electrodes ($\Delta \mu$ is the
corresponding chemical potential difference) so that the photocurrent
works against this bias. The kinetic rates now satisfy
\begin{align}
\label{eq:DBC_2-level_sun}
\frac{k_{10}}{k_{01}} &= e^{-\beta (E_1-\mu)}\, ; \quad \frac{k_{21}}{k_{12}} = e^{-\beta_{\rm\scriptscriptstyle S} (E_2-E_1)}\, ; \quad
\frac{k_{20}}{k_{02}} = e^{-\beta (E_2-\mu)}\, ,
\end{align}
where $\mu_2=\mu_1+\Delta \mu$ and $T=1/k_{\rm\scriptscriptstyle B}\beta$ is the ambient temperature. At steady state, the
current $J$ is the same on all segments of the graph of Fig.~\ref{fig:fig1}(b)
\begin{align}
\label{eq:J_av_2-level}
J&=k_{10} P_0 - k_{01} P_1 = k_{21} P_1 - k_{12} P_2 = k_{02} P_2 - k_{20} P_0
\end{align}
The open circuit (OC) voltage is the bias for which this current
vanish. The existence of such a state again implies that these rates
satisfy Eq.~(\ref{eq:path_ratio_2-level}).
Equations~(\ref{eq:path_ratio_2-level}) and (\ref{eq:DBC_2-level_sun})
then lead to
\begin{align}
\label{eq:Carnot_eff}
\frac{\Delta \mu^{\rm OC}}{E_2-E_1} &= 1-\frac{T}{T_{\rm\scriptscriptstyle S}}
\end{align}
Viewed as the zero current limit of the efficiency $J \Delta
\mu/[J(E_2-E_1)]$ (ratio between the work per unit time, $\dot{W}= J
\Delta \mu$ extracted from the device and the heat per unit time,
$\dot{Q}= (E_2-E_1)J$ absorbed from sun), Eq.~(\ref{eq:Carnot_eff})
simply identifies the efficiency in this reversible (zero current)
limit as the Carnot efficiency. Remarkably, this result does not
depend on the relative alignment of the molecular levels with respect
to the electrodes Fermi levels. It does rely on the assumption that
all input ``sun heat'' enters at the resonance energy $E_2-E_1$, and
identifies the inability of this system to efficiently extract energy
from photons of different energies as an important source of loss.

This simple example demonstrates the use of kinetic schemes that
incorporate rate information in the analysis of photovoltaic device
performance, as well as its relationship to thermodynamics. Naturally,
Carnot efficiency is realized in the OC limit. In the following two
sections we apply a similar analysis to a simple model of bulk
heterojunction organic photovoltaic (BHJ-OPV) cell, where essential
internal losses leads to a maximum efficiency that is lower than the
Carnot result.

\section{BHJ-OPV Model}
\label{sec:BHJ-OPV_model}
\begin{figure}[h!]
  \centering
\includegraphics[width=0.55\textwidth]{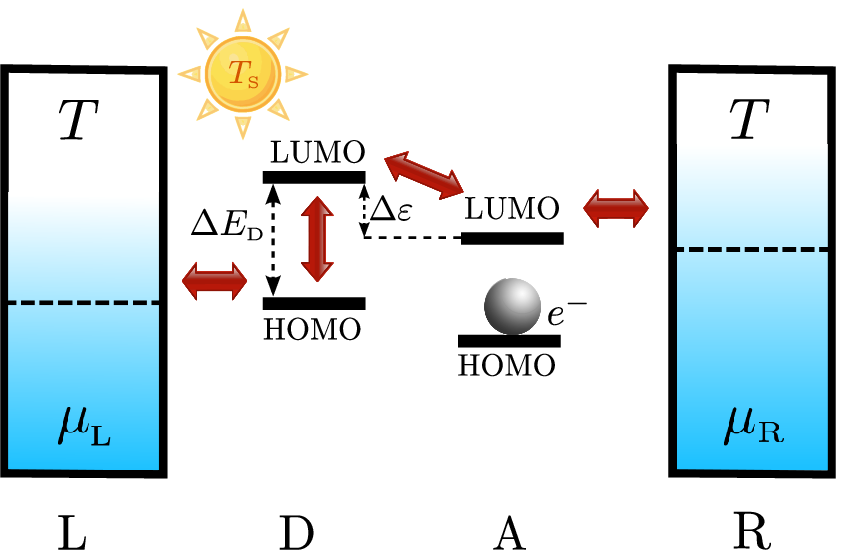}
 \caption{Schematic representation of energetics in BHJ solar cells. The
   system consists of a donor and acceptor, each characterized by
   their HOMO and LUMO levels.}
 \label{fig:fig2}
\end{figure}

The BHJ-OPV cell model considered here is comprised of two effective sites
$l=\rm D,\,A$ representing the donor (D) and the acceptor (A) molecules, in contact with two
electrodes, $L$ and $R$ (see Fig.~\ref{fig:fig2}). Each of the sites is described as a two-state system
with energy levels $\varepsilon_{\rm\scriptscriptstyle  D1},\varepsilon_{\rm\scriptscriptstyle D2}$) and
($\varepsilon_{\rm\scriptscriptstyle A1},\varepsilon_{\rm\scriptscriptstyle A2}$) corresponding to the highest occupied and
lowest unoccupied molecular orbitals (HOMO, LUMO) levels of the donor and acceptor
species, respectively. The electrodes are represented by free-electron reservoirs at
chemical potentials $\mu_{K}$ ($K=L, R$) that are set to $\varepsilon_{\rm\scriptscriptstyle F} =
\varepsilon_{\rm\scriptscriptstyle D1} + \Delta E_{\rm\scriptscriptstyle D}/2$
($\Delta E_{\rm\scriptscriptstyle D} = \varepsilon_{\rm\scriptscriptstyle D2}-\varepsilon_{\rm\scriptscriptstyle D1}$)
in the zero-bias junction. The electrochemical potential difference corresponds to a bias
The electrochemical potential difference corresponds to a bias voltage
$U=(\mu_{\rm\scriptscriptstyle R} - \mu_{\rm\scriptscriptstyle L})/|e|$ where
$|e|$ is the electron charge. In what follows we use the notation $\Delta E_l = \varepsilon_{l2} -
\varepsilon_{l1}$ ($l=\rm D,\,A$), for the energy differences that represent the donor and
acceptor band gaps, and refer to $\Delta\varepsilon = \varepsilon_{\rm\scriptscriptstyle D2} -
\varepsilon_{\rm\scriptscriptstyle A2}$ as the interface or
donor-acceptor LUMO-LUMO gap.\cite{note:EN_2} The different system states are described
by occupation numbers $n_{\rm\scriptscriptstyle K_j}=0,1$, where $K=D,A$ and
$j=1,2$.

\begin{figure}[b!]
\centering
\includegraphics[width=0.55 \textwidth]{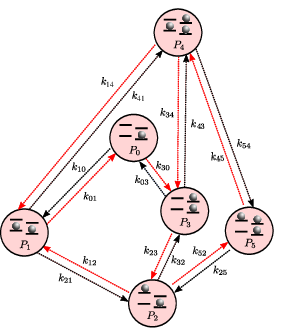}
\caption{(Network representation of
   the underlying master equation associated with the six accessible
   microstates. The graph is composed of six vertices (shown as
   circles). The interconnected vertices represent the probabilities
   $P_N$ to find the system in a microstate $N$ ($N=0,...,5$) and the
   edges connecting some pairs of vertices stand for transitions
   between the states. The edges are drawn as arrows that indicate
   transitions with rate $k_{N'\,N} = k_{N' \leftarrow N}$ from a
   state (vertex) $N$ to $N'$.}
 \label{fig:fig3}
\end{figure}

To further assign realistic contents to this model we introduce the restrictions
$n_{\rm\scriptscriptstyle D1} n_{\rm\scriptscriptstyle  D2} = 0$ (i.e., the donor cannot be double occupied)
and $n_{\rm\scriptscriptstyle A1} =1$. The second condition
implies that the acceptor can only receive (and subsequently release) an additional
electron. Because of this restriction, the energy $\varepsilon_{\rm\scriptscriptstyle A2}$
can be taken as the corresponding single electron energy given that level $A_1$ is occupied.
The resulting microscopic description then consists of six states with respect to the
occupations  $(n_{\rm\scriptscriptstyle
  D1},\,n_{\rm\scriptscriptstyle D2},\,n_{\rm\scriptscriptstyle
  A1},\,n_{\rm\scriptscriptstyle A2})$, that we denote by the integers
$N=0,\,...,\,5$, (see Fig.~\ref{fig:fig3}). Within this six-state
representation, the probability to find the system in state $N$ is denoted by $P_N$. The
system dynamics is modeled by a master equation accounting for the time evolution of
the probabilities $P_N (t)$ ($N=0,\,...,\,5$) fulfilling normalization $\sum_{N} P_N (t) =1$
at all times (for details, see Ref.~\onlinecite{Einax/etal:2011}).
The steady state is evaluated by setting $d P_N (t)/dt =0$.

In what follows, we assume that the transition rates $k_{N'\,N} = k_{N' \leftarrow N}$  from state $N$
to state $N'$ obey (local) detailed balance condition, i.~e. their ratios are given by
$k_{N'\, N}/k_{N\, N'} = \exp(-\beta_\nu \Delta E_{N'\, N})$, where $\Delta E_{N'\, N} = E_{N'}-E_{N}$.
Note that in general $\Delta E_{N'\, N}$ is
determined by intrinsic energy differences as well as external driving forces.\cite{Einax/etal:2010a}
$\beta_\nu=1/k_{\rm\scriptscriptstyle B} T_\nu$ is the inverse thermal energy associated with a thermal bath at temperature
$T_\nu$. As in the $2$-level example addressed in Section~\ref{sec:2-level}, some of the rates processes are
governed by the ambient temperature $T$, while others reflect external driving force. In the
present model the latter are the $1 \rightleftharpoons 2$ and $4  \rightleftharpoons 5$ transitions,
which are governed by the effective temperature $T_{\rm\scriptscriptstyle eff}$ defined below.

Because the heterojunction architecture entails an intrinsic energy loss associated
with the exciton dissociation, the energetics is determined by both the interfacial gap
energy $\Delta \varepsilon$ and the exciton binding energy. In what follows we will define the exciton
binding energy $V_{\rm\scriptscriptstyle C}$ as the difference between the energy needed to move the electron from
donor upper level $D2$ to the acceptor level $A2$, and the same energy evaluated in the
fictitious case in which the Coulombic electron-hole interaction is disregarded. It is
important to note the difference between this many body energy and the essentially single
electron energies $\varepsilon_{\rm\scriptscriptstyle K_j}$, $K=D,A$, $j=1,2$.
The latter are properties of the single electron levels depicted in Fig.~\ref{fig:fig2},
and their differences enter in evaluating the transition energies between the corresponding levels.
In contrast, $V_{\rm\scriptscriptstyle C}$ is a property of a transition between states $2$ and $3$ (see Fig.~\ref{fig:fig3})
and does not enter any other transition energy. (This assumes, as
we do here, that the exciton binding energy is fully realized in this transition, i.~e., that the
corresponding electron-hole Coulomb attraction does not extend beyond the nearest
neighbor $D$-$A$ distance). Thus $E_3-E_2=\varepsilon_{\rm\scriptscriptstyle A2}-\varepsilon_{\rm\scriptscriptstyle D2}+V_{\rm\scriptscriptstyle C}$,
however (for example) $E_1-E_0=E_4-E_3=\varepsilon_{\rm\scriptscriptstyle D1}-\mu_L$
and $E_3-E_0=E_5-E_2=\varepsilon_{\rm\scriptscriptstyle A2}-\mu_R$ do not depend on $V_{\rm\scriptscriptstyle C}$.
With this understanding, the energy differences $\Delta E_{N'\, N} = E_{N'}-E_{N}$ between any two molecular states
depicted in Fig.~\ref{fig:fig3} can be written and used as described below.

\section{Thermodynamic efficiency limit from a cycle representations}
\label{sec:thermo_eff_limit_CP}
The six system states shown in Fig.~\ref{fig:fig3} are connected by rate processes, forming
a graph in which the states are represented by nodes while the rate processes corresponds
to the links between them. This graph can be decomposed into cycles, as detailed in
Table~\ref{table:cycle}.

Let us focus on the fundamental cycle associated with the path
$C_1$: $P_0 \rightarrow P_1 \rightarrow P_2 \rightarrow P_3 \rightarrow P_0$.
This cycle represents the photovoltaic operation of the
considered minimal model for a BHJ-OPV solar cell. In the ``forward direction'' it starts
with electron transfer from the left electrode into level $D1$ ($P_0 \rightarrow P_1$),
followed by light induced promotion of the electron to level $D2$ ($P_1 \rightarrow P_2$),
exciton dissociation, that is electron transfer from $D2$ to $A2$ ($P_2 \rightarrow P_3$) and,
finally, transfer of the excess electron on level $A2$ of the acceptor to the right electrode.
These processes are of course accompanied by their reverse counterparts.
The energies associated with these transitions are
$\Delta E_{10} = \varepsilon_{\rm\scriptscriptstyle D1}-\mu_{\rm\scriptscriptstyle L}$,
$\Delta E_{21} = \varepsilon_{\rm\scriptscriptstyle D2}-\varepsilon_{\rm\scriptscriptstyle D1} = \Delta E_D$,
$\Delta E_{32} = \varepsilon_{\rm\scriptscriptstyle A2}-\varepsilon_{\rm\scriptscriptstyle D2} + V_{\rm\scriptscriptstyle C} = V_{\rm\scriptscriptstyle C} - \Delta \varepsilon$,
and $\Delta E_{03} = \mu_{\rm\scriptscriptstyle R} -\varepsilon_{\rm\scriptscriptstyle A2}$.
The corresponding rates satisfy detailed balance conditions that are
determined by these energies and the corresponding temperatures. The processes $0 \rightleftharpoons 1$,
$2 \rightleftharpoons 3$, and $3 \rightleftharpoons 0$
are governed by the ambient temperature $T$. Consequently
\begin{align}
\label{eq:k10k01}
\frac{k_{10}}{k_{01}} &= e^{-(\varepsilon_{\rm\scriptscriptstyle D1}-\mu_{\rm\scriptscriptstyle L})/k_{\rm\scriptscriptstyle B} T} \, , \\
\label{eq:k32k23}
\frac{k_{32}}{k_{23}} &= e^{-(V_{\rm\scriptscriptstyle C} - \Delta \varepsilon)/k_{\rm\scriptscriptstyle B} T} \, ,
\end{align}
and
\begin{align}
\label{eq:k03k30}
\frac{k_{03}}{k_{30}} &= e^{-(\mu_{\rm\scriptscriptstyle R}-{\varepsilon}_{\rm\scriptscriptstyle A2})/k_{\rm\scriptscriptstyle B} T} \, .
\end{align}
Consider now the photoinduced $1 \rightleftharpoons 2$ process. In general, both the forward and
reverse transitions are associated with radiative and non-radiative excitation and
recombination
\begin{align}
\label{eq:k21k12_sum}
k_{12} &= k_{12}^{\rm R} + k_{12}^{\rm NR}\, ; \quad k_{21} = k_{21}^{\rm R} + k_{21}^{\rm NR} \, .
\end{align}
The radiative rates are photoinduced by sunlight and satisfy a detailed balance condition
associated with the sun temperature $T_{\rm\scriptscriptstyle S}$, while the non- radiative rates are determined by
interaction with the environment and obey a detailed balance relation governed by the
ambient temperature
\begin{align}
\label{eq:k21k12}
\frac{k_{21}^{R}}{k_{12}^{R}} &= e^{-\Delta E_{\rm\scriptscriptstyle D} /k_{\rm\scriptscriptstyle B} T_{\rm\scriptscriptstyle S} }\, ; \quad
\frac{k_{21}^{NR}}{k_{12}^{NR}} = e^{-\Delta E_{\rm\scriptscriptstyle D} /k_{\rm\scriptscriptstyle B} T }\, .
\end{align}
\begin{table}
  \caption{Cycles associated with the network of the systems states given in Fig.~\ref{fig:fig2}.}
  \vspace*{1ex}
  \label{table:cycle}
  \begin{tabular}{|c|l|}
    \hline
    & \\[-1ex]
    CYCLE & \hspace*{2cm} PATH \\[1ex]
    \hline \hline
    & \\[-1ex]
    C$_1$ & $P_0 \rightarrow P_1 \rightarrow P_2 \rightarrow P_3 \rightarrow P_0$\\[2ex]
    C$_2$ & $P_0 \rightarrow P_1 \rightarrow P_4 \rightarrow P_5 \rightarrow P_2 \rightarrow P_3 \rightarrow P_0$\\[2ex]
    C$_3$ & $P_1 \rightarrow P_2 \rightarrow P_3 \rightarrow P_4 \rightarrow P_1$\\[2ex]
    $C_4$ & $P_2 \rightarrow P_3 \rightarrow P_4 \rightarrow P_5 \rightarrow P_2$ \\[2ex]
    C$_5$ & $P_1 \rightarrow P_2 \rightarrow P_5 \rightarrow P_4 \rightarrow P_1$\\[2ex]
    C$_6$ & $P_0 \rightarrow P_1 \rightarrow P_4 \rightarrow P_3 \rightarrow P_0$\\[1ex]
    \hline
  \end{tabular}
\end{table}
Consequently
\begin{align}
\label{eq:k21k12_tot}
\frac{k_{21}}{k_{12}} &\equiv \frac{k_{21}^{\rm R} + k_{21}^{\rm NR} }{k_{12}^{\rm R} + k_{12}^{\rm NR}} =  e^{-\Delta E_{\rm\scriptscriptstyle D} /k_{\rm\scriptscriptstyle B} T_{\rm\scriptscriptstyle eff} } \, ,
\end{align}
where the effective temperature $T_{\rm\scriptscriptstyle eff}$ is defined by
 \begin{align}
\label{eq:Temperatur_eff}
T_{\rm\scriptscriptstyle eff} &= \frac{\Delta E_{\rm\scriptscriptstyle D}}{k_{\rm\scriptscriptstyle B}} \frac{1}{\ln \big[ \frac{k_{12}^{\rm R} + k_{12}^{\rm NR} }{k_{21}^{\rm R} + k_{21}^{\rm NR}}\big]} \, .
\end{align}
In the absence of radiationless loss ($k_{12}^{\rm NR}=k_{21}^{\rm NR}=0$) $T_{\rm\scriptscriptstyle eff}=T_{\rm\scriptscriptstyle S}$.
In the presence of such loss, Eq.~(\ref{eq:Temperatur_eff}) implies that (since $T < T_{\rm\scriptscriptstyle S}$)
$T_{\rm\scriptscriptstyle eff} < T_{\rm\scriptscriptstyle S}$ . Note that the absolute magnitude of
$T_{\rm\scriptscriptstyle eff}$ is determined not only by the temperatures $T$ and $T_{\rm\scriptscriptstyle eff}$ but also by the kinetic rates
themselves: faster non-radiative recombination implies lower effective temperature.

Next, suppose that the cycle $C_1$ represents the entire energy conversion device.
Consider the ratio of products of forward and backward, rates, $k_{10} k_{21} k_{32} k_{03}$ and $k_{01} k_{12} k_{23} k_{30}$ in cycle $C_1$.
From Eqs.~(\ref{eq:k10k01})-(\ref{eq:k03k30}) and (\ref{eq:k21k12_tot}) we get
\begin{align}
\label{eq:ratio_C1}
\frac{k_{10} k_{21} k_{32} k_{03}}{ k_{01} k_{12} k_{23} k_{30}} &= e^{- (\Delta \mu  - \Delta E_{\rm\scriptscriptstyle D} + V_{\rm\scriptscriptstyle C})/k_{\rm\scriptscriptstyle B} T} e^{- \Delta E_{\rm\scriptscriptstyle D}/k_{\rm\scriptscriptstyle B} T_{\rm\scriptscriptstyle eff}} \equiv
e^{-A(C_1)/k_{\rm\scriptscriptstyle B} T}\, ,
\end{align}
where $\Delta \mu = \mu_{\rm\scriptscriptstyle R}-\mu_{\rm\scriptscriptstyle L}$.
The quantity $A(C_1)$ defined by (\ref{eq:ratio_C1}) is the affinity of the cycle $C_1$.
It can be recast in the form
\begin{align}
\label{eq:cycle_affinity}
A(C_1) &=  \frac{ \Delta \mu  + V_{\rm\scriptscriptstyle C} - \Delta E_{\rm\scriptscriptstyle D} \eta_{\rm\scriptscriptstyle eff}^{\rm\scriptscriptstyle C} }{
k_{\rm\scriptscriptstyle B} T} \, ,
\end{align}
where
\begin{align}
\label{eq:eta_eff}
\eta_{\rm\scriptscriptstyle eff}^{\rm\scriptscriptstyle C} &= 1-\frac{T}{T_{\rm\scriptscriptstyle eff}}
\end{align}
is the Carnot efficiency of a reversible machine operating between temperatures $T$ and $T_{\rm\scriptscriptstyle eff}$.
As discussed in Sec.~\ref{sec:2-level}, the cycle affinity vanishes when the cycle carries no current.
In this reversible case Eq.~(\ref{eq:cycle_affinity}) yields
\begin{align}
\label{eq:OCV}
\frac{\Delta \mu^{\rm OC}}{\Delta E_{\rm\scriptscriptstyle D}} &= \eta_{\rm\scriptscriptstyle eff}^{\rm\scriptscriptstyle C} -
\frac{V_{\rm\scriptscriptstyle C}}{\Delta E_{\rm\scriptscriptstyle D}} \, .
\end{align}
As discussed in Sec.~\ref{sec:2-level} [see Eq.~(\ref{eq:Carnot_eff})], the left hand side of this equation represents the
energy conversion efficiency of our device. When $T_{\rm\scriptscriptstyle eff}=T_{\rm\scriptscriptstyle S}$
(i.~e. in the absence of nonradiative recombination) and $V_{\rm\scriptscriptstyle C} = 0$ (vanishing exciton binding energy),
this device operates, in this open circuit limit, at the Carnot efficiency associated with the sun
temperature. Equation~(\ref{eq:OCV}) shows explicitly the two sources of efficiency reduction in this
reversible (open voltage) situation: The presence of non-radiative recombination which
renders an effective temperature lower than $T_{\rm\scriptscriptstyle S}$ and the exciton binding energy that needs
to be overcome during the operation at the cost of useful work.

The result (\ref{eq:OCV}) is an expression for the maximal efficiency of a device operating
along cycle $C_1$. However, it is easily checked that the same condition for vanishing
affinity is obtained for any of the cycles in Table~\ref{table:cycle} that contains the exciton dissociation
($2 \rightleftharpoons 3$) step, namely cycles $C_1$,$C_2$,$C_3$, and $C_4$.
(To verify this note that $k_{43}/k_{34}= k_{10}/k_{01}$,
and $k_{14}/k_{41}=k_{25}/k_{52}=k_{03}/k_{30}$). Furthermore, for the both cycles $C_5$ and $C_6$ we find
$A(C_5)=A(C_6) = 0$. Therefore the result (\ref{eq:OCV}) is valid for the original $6$-state system
depicted in Figs.~\ref{fig:fig2} and \ref{fig:fig3}. Note that in the absence of non-radiative
recombination, Eq.~(\ref{eq:OCV}) becomes
\begin{align}
\label{eq:OCV_ideal}
\frac{\Delta \mu^{\rm OC}}{\Delta E_{\rm\scriptscriptstyle D}} &= \eta^{\rm\scriptscriptstyle C} -
\frac{V_{\rm\scriptscriptstyle C}}{\Delta E_{\rm\scriptscriptstyle D}} \, ; \quad \eta^{\rm\scriptscriptstyle C} =
1 - \frac{T}{T_{\rm\scriptscriptstyle S}} \, ,
\end{align}
which is compatible with the result of Ref.~\onlinecite{Giebink/etal:2011}.

Equations~(\ref{eq:Temperatur_eff}), (\ref{eq:eta_eff}), and (\ref{eq:OCV}) provide a simple and transparent view of the sources of
OC voltage reduction and reversible efficiency loss in BHJ-OPV cells. We have checked
this result by solving the underlying master equation given in\cite{Einax/etal:2010a}.
To this end we have adapted the energetics and the transitions rates used in this previously work:\cite{Einax/etal:2010a}
$\mu_{\rm\scriptscriptstyle L}=0.0\, \rm eV$,
$\mu_{\rm\scriptscriptstyle R}=\mu_{\rm\scriptscriptstyle L} + \Delta \mu$,
$\varepsilon_{\rm\scriptscriptstyle D1} = -0.1\,\rm eV$,
$\varepsilon_{\rm\scriptscriptstyle D2} = 1.4\,\rm eV$,
$\varepsilon_{\rm\scriptscriptstyle A2} = 1.15\,\rm eV$ and
$V_{\rm\scriptscriptstyle C} = 0.15\,\rm eV$, and have set
the temperatures to $T=300 \rm K$ and $T_{\rm\scriptscriptstyle S}=6000 \rm K$
so the Carnot efficiency is $\eta^{\rm\scriptscriptstyle C}=0.95$.
For simplicity we neglect radiationless losses on the donor.\cite{note:EN_3}
The numerical calculation gives the open circuit voltage $\Delta \mu^{\rm OC} = 1.275$\, eV,
which agrees exactly with that value predicted by Eq.~(\ref{eq:OCV_ideal}), i.~e.,
$\Delta\mu^{\rm OC} = 0.95 \Delta E_{\rm\scriptscriptstyle D} - 0.15\,{\rm eV} =1.275$\, eV.
For $V_{\rm\scriptscriptstyle C} = 0.15$\, eV, the maximal achievable thermodynamic efficiency
is $\eta^{\rm\scriptscriptstyle th}= 0.85$.

\section{Conclusion and Perspectives}
\label{sec:summary}
We have presented a novel concept for performance analysis of photovoltaic cells
and have applied it to the simplest $2$-level device model as well as a generic model for an
organic photovoltaic cell. The starting point is the modelling of the energy conversion
process by a set of kinetic (master) equations with rate coefficients that incorporate the
system energy level structure as well as the relevant energetic, thermal and optical
constraints and driving forces. Further analysis is facilitated by describing the resulting
master equation as a graph in which the rates are represented by edges that link between
vortices representing states. This makes it possible to exploit the decomposition of the
network into cycles to get better insight on the interrelations between the physical fluxes.
Such a kinetic scheme can be used to analyze the system performance at and away from
equilibrium, however in this paper we have focused on open circuit (OC) situations, in
particular the simplest subclass of those in which all internal currents, therefore all cycle
affinities, vanish. The performance of such systems does not depends on individual rates,
only on ratios between backward and forward rates that are determined by detailed
balance conditions. For the $2$-level/$3$-state device model of references 30 and 32 this
analysis yields the Carnot value for the maximum OC efficiency. A similar calculation
for a generic model of a bulk heterojunction organic photovoltaic (BHJ-OPV) cell that
incorporates the exciton dissociation energy as well as non-radiative recombination in the
donor-subsystem leads to a maximum OC efficiency and OC voltage that are lower than
the limiting Carnot value. For example, with our choice of (reasonable) parameters the
maximum available efficiency is found to be $0.85$ , which $\sim10\%$ lower than the
corresponding Carnot value ($\sim 0.95$). This approach can be generalized in several ways.
Operation under finite overall current can be analyzed to yield efficiency at maximum
power.\cite{Einax/Nitzan:2014} Even under OC conditions, loss due to the presence of cycles with nonvanishing
currents can be encountered in more complex models and should be accounted
for. Finally, extending such approach to the quantum-mechanical regime may be of
interest. These will be subjects of future efforts.

\begin{acknowledgements}
  The research is supported by the Israel Science Foundation, the
  Israel-US Binational Science Foundation (grant No. 2011509), and the European Science
  Council (FP7/ERC grant No. 226628). We thank Mark Ratner, Philip Ruyten and Bart
  Cleuren for stimulating discussions. AN thanks the Chemistry Department at the
  University of Pennsylvania for hospitality during the time this paper was completed.
\end{acknowledgements}

\bibliographystyle{apsrev4-1}
\bibliography{OPV_cells}

\begin{thebibliography}{60}%
\makeatletter
\providecommand \@ifxundefined [1]{%
 \@ifx{#1\undefined}
}%
\providecommand \@ifnum [1]{%
 \ifnum #1\expandafter \@firstoftwo
 \else \expandafter \@secondoftwo
 \fi
}%
\providecommand \@ifx [1]{%
 \ifx #1\expandafter \@firstoftwo
 \else \expandafter \@secondoftwo
 \fi
}%
\providecommand \natexlab [1]{#1}%
\providecommand \enquote  [1]{``#1''}%
\providecommand \bibnamefont  [1]{#1}%
\providecommand \bibfnamefont [1]{#1}%
\providecommand \citenamefont [1]{#1}%
\providecommand \href@noop [0]{\@secondoftwo}%
\providecommand \href [0]{\begingroup \@sanitize@url \@href}%
\providecommand \@href[1]{\@@startlink{#1}\@@href}%
\providecommand \@@href[1]{\endgroup#1\@@endlink}%
\providecommand \@sanitize@url [0]{\catcode `\\12\catcode `\$12\catcode
  `\&12\catcode `\#12\catcode `\^12\catcode `\_12\catcode `\%12\relax}%
\providecommand \@@startlink[1]{}%
\providecommand \@@endlink[0]{}%
\providecommand \url  [0]{\begingroup\@sanitize@url \@url }%
\providecommand \@url [1]{\endgroup\@href {#1}{\urlprefix }}%
\providecommand \urlprefix  [0]{URL }%
\providecommand \Eprint [0]{\href }%
\providecommand \doibase [0]{http://dx.doi.org/}%
\providecommand \selectlanguage [0]{\@gobble}%
\providecommand \bibinfo  [0]{\@secondoftwo}%
\providecommand \bibfield  [0]{\@secondoftwo}%
\providecommand \translation [1]{[#1]}%
\providecommand \BibitemOpen [0]{}%
\providecommand \bibitemStop [0]{}%
\providecommand \bibitemNoStop [0]{.\EOS\space}%
\providecommand \EOS [0]{\spacefactor3000\relax}%
\providecommand \BibitemShut  [1]{\csname bibitem#1\endcsname}%
\let\auto@bib@innerbib\@empty
\bibitem [{\citenamefont {Nelson}(2003)}]{Nelson:2003}%
  \BibitemOpen
  \bibfield  {author} {\bibinfo {author} {\bibfnamefont {J.}~\bibnamefont
  {Nelson}},\ }\href@noop {} {\emph {\bibinfo {title} {The Physics of Solar
  Cells}}}\ (\bibinfo  {publisher} {World Scientific},\ \bibinfo {address}
  {Singapore},\ \bibinfo {year} {2003})\BibitemShut {NoStop}%
\bibitem [{\citenamefont {W{\"u}rfel}(2009)}]{Wuerfel:2009}%
  \BibitemOpen
  \bibfield  {author} {\bibinfo {author} {\bibfnamefont {P.}~\bibnamefont
  {W{\"u}rfel}},\ }\href@noop {} {\emph {\bibinfo {title} {Physics of Solar
  Cells: From Basic Principles to Advanced Concepts}}},\ \bibinfo {edition}
  {2nd}\ ed.\ (\bibinfo  {publisher} {Wiley VCH},\ \bibinfo {address}
  {Weinheim},\ \bibinfo {year} {2009})\BibitemShut {NoStop}%
\bibitem [{\citenamefont {Nayak}\ \emph {et~al.}(2012)\citenamefont {Nayak},
  \citenamefont {Garcia-Belmonte}, \citenamefont {Kahn}, \citenamefont
  {Bisquert},\ and\ \citenamefont {Cahen}}]{Nayak/etal:2012}%
  \BibitemOpen
  \bibfield  {author} {\bibinfo {author} {\bibfnamefont {P.~K.}\ \bibnamefont
  {Nayak}}, \bibinfo {author} {\bibfnamefont {G.}~\bibnamefont
  {Garcia-Belmonte}}, \bibinfo {author} {\bibfnamefont {A.}~\bibnamefont
  {Kahn}}, \bibinfo {author} {\bibfnamefont {J.}~\bibnamefont {Bisquert}}, \
  and\ \bibinfo {author} {\bibfnamefont {D.}~\bibnamefont {Cahen}},\
  }\href@noop {} {\bibfield  {journal} {\bibinfo  {journal} {Energy Environ.
  Sci.}\ }\textbf {\bibinfo {volume} {5}},\ \bibinfo {pages} {6022} (\bibinfo
  {year} {2012})}\BibitemShut {NoStop}%
\bibitem [{\citenamefont {Br{\"u}tting}\ and\ \citenamefont
  {Adachi}(2012)}]{Bruetting:2012}%
  \BibitemOpen
  \bibfield  {author} {\bibinfo {author} {\bibfnamefont {W.}~\bibnamefont
  {Br{\"u}tting}}\ and\ \bibinfo {author} {\bibfnamefont {C.}~\bibnamefont
  {Adachi}},\ }\href@noop {} {\emph {\bibinfo {title} {Physics of Organic
  Semiconductors}}},\ \bibinfo {edition} {2nd}\ ed.\ (\bibinfo  {publisher}
  {Wiley VCH},\ \bibinfo {address} {Weinheim},\ \bibinfo {year}
  {2012})\BibitemShut {NoStop}%
\bibitem [{\citenamefont {Dennler}\ \emph {et~al.}(2009)\citenamefont
  {Dennler}, \citenamefont {Scharber},\ and\ \citenamefont
  {Brabec}}]{Denner/etal:2009}%
  \BibitemOpen
  \bibfield  {author} {\bibinfo {author} {\bibfnamefont {G.}~\bibnamefont
  {Dennler}}, \bibinfo {author} {\bibfnamefont {M.~C.}\ \bibnamefont
  {Scharber}}, \ and\ \bibinfo {author} {\bibfnamefont {C.~J.}\ \bibnamefont
  {Brabec}},\ }\href@noop {} {\bibfield  {journal} {\bibinfo  {journal} {Adv.
  Mater.}\ }\textbf {\bibinfo {volume} {21}},\ \bibinfo {pages} {1323}
  (\bibinfo {year} {2009})}\BibitemShut {NoStop}%
\bibitem [{\citenamefont {Chen}\ \emph {et~al.}(2009)\citenamefont {Chen},
  \citenamefont {Hou}, \citenamefont {Zhang}, \citenamefont {Liang},
  \citenamefont {Yang}, \citenamefont {Yang}, \citenamefont {Yu}, \citenamefont
  {Wu},\ and\ \citenamefont {Li}}]{Chen/etal:2009}%
  \BibitemOpen
  \bibfield  {author} {\bibinfo {author} {\bibfnamefont {H.-Y.}\ \bibnamefont
  {Chen}}, \bibinfo {author} {\bibfnamefont {J.}~\bibnamefont {Hou}}, \bibinfo
  {author} {\bibfnamefont {S.}~\bibnamefont {Zhang}}, \bibinfo {author}
  {\bibfnamefont {Y.}~\bibnamefont {Liang}}, \bibinfo {author} {\bibfnamefont
  {G.}~\bibnamefont {Yang}}, \bibinfo {author} {\bibfnamefont {Y.}~\bibnamefont
  {Yang}}, \bibinfo {author} {\bibfnamefont {L.}~\bibnamefont {Yu}}, \bibinfo
  {author} {\bibfnamefont {Y.}~\bibnamefont {Wu}}, \ and\ \bibinfo {author}
  {\bibfnamefont {G.}~\bibnamefont {Li}},\ }\href@noop {} {\bibfield  {journal}
  {\bibinfo  {journal} {Nature Photonics}\ }\textbf {\bibinfo {volume} {3}},\
  \bibinfo {pages} {649} (\bibinfo {year} {2009})}\BibitemShut {NoStop}%
\bibitem [{\citenamefont {Bredas}\ \emph {et~al.}(2009)\citenamefont {Bredas},
  \citenamefont {Norton}, \citenamefont {Cornil},\ and\ \citenamefont
  {Coropceanu}}]{Bredas/etal:2009}%
  \BibitemOpen
  \bibfield  {author} {\bibinfo {author} {\bibfnamefont {J.-L.}\ \bibnamefont
  {Bredas}}, \bibinfo {author} {\bibfnamefont {J.~E.}\ \bibnamefont {Norton}},
  \bibinfo {author} {\bibfnamefont {J.}~\bibnamefont {Cornil}}, \ and\ \bibinfo
  {author} {\bibfnamefont {V.}~\bibnamefont {Coropceanu}},\ }\href@noop {}
  {\bibfield  {journal} {\bibinfo  {journal} {Accounts of Chemical Research}\
  }\textbf {\bibinfo {volume} {42}},\ \bibinfo {pages} {1691} (\bibinfo {year}
  {2009})}\BibitemShut {NoStop}%
\bibitem [{\citenamefont {Deibel}\ and\ \citenamefont
  {Dyakonov}(2010)}]{Deibel/Dyakonov:2010}%
  \BibitemOpen
  \bibfield  {author} {\bibinfo {author} {\bibfnamefont {C.}~\bibnamefont
  {Deibel}}\ and\ \bibinfo {author} {\bibfnamefont {V.}~\bibnamefont
  {Dyakonov}},\ }\href@noop {} {\bibfield  {journal} {\bibinfo  {journal} {Rep.
  Prog. Phys.}\ }\textbf {\bibinfo {volume} {73}},\ \bibinfo {pages} {096401}
  (\bibinfo {year} {2010})}\BibitemShut {NoStop}%
\bibitem [{\citenamefont {Nicholson}\ and\ \citenamefont
  {Castro}(2010)}]{Nicholson/Castro:2010}%
  \BibitemOpen
  \bibfield  {author} {\bibinfo {author} {\bibfnamefont {P.~G.}\ \bibnamefont
  {Nicholson}}\ and\ \bibinfo {author} {\bibfnamefont {F.~A.}\ \bibnamefont
  {Castro}},\ }\href@noop {} {\bibfield  {journal} {\bibinfo  {journal}
  {Nanotechnology}\ }\textbf {\bibinfo {volume} {21}},\ \bibinfo {pages}
  {492001} (\bibinfo {year} {2010})}\BibitemShut {NoStop}%
\bibitem [{\citenamefont {Thompson}\ \emph {et~al.}(2011)\citenamefont
  {Thompson}, \citenamefont {Khlyabich}, \citenamefont {Burkhart},
  \citenamefont {Aviles}, \citenamefont {Rudenko}, \citenamefont {Shultz},
  \citenamefont {Ng},\ and\ \citenamefont {Mangubat}}]{Thompson/etal:2011}%
  \BibitemOpen
  \bibfield  {author} {\bibinfo {author} {\bibfnamefont {B.~C.}\ \bibnamefont
  {Thompson}}, \bibinfo {author} {\bibfnamefont {P.~P.}\ \bibnamefont
  {Khlyabich}}, \bibinfo {author} {\bibfnamefont {B.}~\bibnamefont {Burkhart}},
  \bibinfo {author} {\bibfnamefont {A.~E.}\ \bibnamefont {Aviles}}, \bibinfo
  {author} {\bibfnamefont {A.}~\bibnamefont {Rudenko}}, \bibinfo {author}
  {\bibfnamefont {G.~V.}\ \bibnamefont {Shultz}}, \bibinfo {author}
  {\bibfnamefont {C.~F.}\ \bibnamefont {Ng}}, \ and\ \bibinfo {author}
  {\bibfnamefont {L.~B.}\ \bibnamefont {Mangubat}},\ }\href@noop {} {\bibfield
  {journal} {\bibinfo  {journal} {Green}\ }\textbf {\bibinfo {volume} {1}},\
  \bibinfo {pages} {29} (\bibinfo {year} {2011})}\BibitemShut {NoStop}%
\bibitem [{\citenamefont {Nelson}(2011)}]{Nelson:2011}%
  \BibitemOpen
  \bibfield  {author} {\bibinfo {author} {\bibfnamefont {J.}~\bibnamefont
  {Nelson}},\ }\href@noop {} {\bibfield  {journal} {\bibinfo  {journal}
  {Materials Today}\ }\textbf {\bibinfo {volume} {14}},\ \bibinfo {pages} {462}
  (\bibinfo {year} {2011})}\BibitemShut {NoStop}%
\bibitem [{\citenamefont {Camaioni}\ and\ \citenamefont
  {Po}(2013)}]{Camaioni/Po:2013}%
  \BibitemOpen
  \bibfield  {author} {\bibinfo {author} {\bibfnamefont {N.}~\bibnamefont
  {Camaioni}}\ and\ \bibinfo {author} {\bibfnamefont {R.}~\bibnamefont {Po}},\
  }\href@noop {} {\bibfield  {journal} {\bibinfo  {journal} {The Journal of
  Physical Chemistry Letters}\ }\textbf {\bibinfo {volume} {4}},\ \bibinfo
  {pages} {1821} (\bibinfo {year} {2013})}\BibitemShut {NoStop}%
\bibitem [{\citenamefont {Seki}\ \emph {et~al.}(2013)\citenamefont {Seki},
  \citenamefont {Furube},\ and\ \citenamefont {Yoshida}}]{Seki/etal:2013}%
  \BibitemOpen
  \bibfield  {author} {\bibinfo {author} {\bibfnamefont {K.}~\bibnamefont
  {Seki}}, \bibinfo {author} {\bibfnamefont {A.}~\bibnamefont {Furube}}, \ and\
  \bibinfo {author} {\bibfnamefont {Y.}~\bibnamefont {Yoshida}},\ }\href@noop
  {} {\bibfield  {journal} {\bibinfo  {journal} {Appl. Phys. Lett.}\ }\textbf
  {\bibinfo {volume} {103}},\ \bibinfo {eid} {253904} (\bibinfo {year}
  {2013})}\BibitemShut {NoStop}%
\bibitem [{\citenamefont {Potscavage}\ \emph {et~al.}(2009)\citenamefont
  {Potscavage}, \citenamefont {Sharma},\ and\ \citenamefont
  {Kippelen}}]{Potscavage/etal:2009}%
  \BibitemOpen
  \bibfield  {author} {\bibinfo {author} {\bibfnamefont {W.~J.}\ \bibnamefont
  {Potscavage}}, \bibinfo {author} {\bibfnamefont {A.}~\bibnamefont {Sharma}},
  \ and\ \bibinfo {author} {\bibfnamefont {B.}~\bibnamefont {Kippelen}},\
  }\href@noop {} {\bibfield  {journal} {\bibinfo  {journal} {Acc. Chem. Res.}\
  }\textbf {\bibinfo {volume} {42}},\ \bibinfo {pages} {1758} (\bibinfo {year}
  {2009})}\BibitemShut {NoStop}%
\bibitem [{\citenamefont {Wagenpfahl}\ \emph {et~al.}(2010)\citenamefont
  {Wagenpfahl}, \citenamefont {Deibel},\ and\ \citenamefont
  {Dyakonov}}]{Wagenpfahl/etal:2010}%
  \BibitemOpen
  \bibfield  {author} {\bibinfo {author} {\bibfnamefont {A.}~\bibnamefont
  {Wagenpfahl}}, \bibinfo {author} {\bibfnamefont {C.}~\bibnamefont {Deibel}},
  \ and\ \bibinfo {author} {\bibfnamefont {V.}~\bibnamefont {Dyakonov}},\
  }\href@noop {} {\bibfield  {journal} {\bibinfo  {journal} {IEEE J. Sel. Top.
  Quantum Electron.}\ }\textbf {\bibinfo {volume} {16}},\ \bibinfo {pages}
  {1759} (\bibinfo {year} {2010})}\BibitemShut {NoStop}%
\bibitem [{\citenamefont {Koster}\ \emph {et~al.}(2012)\citenamefont {Koster},
  \citenamefont {Shaheen},\ and\ \citenamefont {Hummelen}}]{Koster/etal:2012}%
  \BibitemOpen
  \bibfield  {author} {\bibinfo {author} {\bibfnamefont {L.~J.~A.}\
  \bibnamefont {Koster}}, \bibinfo {author} {\bibfnamefont {S.~E.}\
  \bibnamefont {Shaheen}}, \ and\ \bibinfo {author} {\bibfnamefont {J.~C.}\
  \bibnamefont {Hummelen}},\ }\href@noop {} {\bibfield  {journal} {\bibinfo
  {journal} {Adv. Energy Mater.}\ }\textbf {\bibinfo {volume} {2}},\ \bibinfo
  {pages} {1246} (\bibinfo {year} {2012})}\BibitemShut {NoStop}%
\bibitem [{\citenamefont {Gruber}\ \emph {et~al.}(2012)\citenamefont {Gruber},
  \citenamefont {Wagner}, \citenamefont {Klein}, \citenamefont {H\"ormann},
  \citenamefont {Opitz}, \citenamefont {Stutzmann},\ and\ \citenamefont
  {Br\"utting}}]{Gruber/etal:2012}%
  \BibitemOpen
  \bibfield  {author} {\bibinfo {author} {\bibfnamefont {M.}~\bibnamefont
  {Gruber}}, \bibinfo {author} {\bibfnamefont {J.}~\bibnamefont {Wagner}},
  \bibinfo {author} {\bibfnamefont {K.}~\bibnamefont {Klein}}, \bibinfo
  {author} {\bibfnamefont {U.}~\bibnamefont {H\"ormann}}, \bibinfo {author}
  {\bibfnamefont {A.}~\bibnamefont {Opitz}}, \bibinfo {author} {\bibfnamefont
  {M.}~\bibnamefont {Stutzmann}}, \ and\ \bibinfo {author} {\bibfnamefont
  {W.}~\bibnamefont {Br\"utting}},\ }\href@noop {} {\bibfield  {journal}
  {\bibinfo  {journal} {Adv. Energy Mater.}\ }\textbf {\bibinfo {volume} {2}},\
  \bibinfo {pages} {1100} (\bibinfo {year} {2012})}\BibitemShut {NoStop}%
\bibitem [{\citenamefont {Shockley}\ and\ \citenamefont
  {Queisser}(1961)}]{Shockley/Queisser:1961}%
  \BibitemOpen
  \bibfield  {author} {\bibinfo {author} {\bibfnamefont {W.}~\bibnamefont
  {Shockley}}\ and\ \bibinfo {author} {\bibfnamefont {H.~J.}\ \bibnamefont
  {Queisser}},\ }\href@noop {} {\bibfield  {journal} {\bibinfo  {journal} {J.
  Appl. Phys.}\ }\textbf {\bibinfo {volume} {32}},\ \bibinfo {pages} {510}
  (\bibinfo {year} {1961})}\BibitemShut {NoStop}%
\bibitem [{\citenamefont {Henry}(1980)}]{Henry:1980}%
  \BibitemOpen
  \bibfield  {author} {\bibinfo {author} {\bibfnamefont {C.~H.}\ \bibnamefont
  {Henry}},\ }\href@noop {} {\bibfield  {journal} {\bibinfo  {journal} {J.
  Appl. Phys.}\ }\textbf {\bibinfo {volume} {51}},\ \bibinfo {pages} {4494}
  (\bibinfo {year} {1980})}\BibitemShut {NoStop}%
\bibitem [{\citenamefont {Landsberg}\ and\ \citenamefont
  {Tonge}(1980)}]{Landsberg/Tonge:1980}%
  \BibitemOpen
  \bibfield  {author} {\bibinfo {author} {\bibfnamefont {P.~T.}\ \bibnamefont
  {Landsberg}}\ and\ \bibinfo {author} {\bibfnamefont {G.}~\bibnamefont
  {Tonge}},\ }\href@noop {} {\bibfield  {journal} {\bibinfo  {journal} {J.
  Appl. Phys.}\ }\textbf {\bibinfo {volume} {51}},\ \bibinfo {pages} {R1}
  (\bibinfo {year} {1980})}\BibitemShut {NoStop}%
\bibitem [{\citenamefont {Giebink}\ \emph {et~al.}(2011)\citenamefont
  {Giebink}, \citenamefont {Wiederrecht}, \citenamefont {Wasieleswski},\ and\
  \citenamefont {Forrest}}]{Giebink/etal:2011}%
  \BibitemOpen
  \bibfield  {author} {\bibinfo {author} {\bibfnamefont {N.~C.}\ \bibnamefont
  {Giebink}}, \bibinfo {author} {\bibfnamefont {G.~P.}\ \bibnamefont
  {Wiederrecht}}, \bibinfo {author} {\bibfnamefont {M.~R.}\ \bibnamefont
  {Wasieleswski}}, \ and\ \bibinfo {author} {\bibfnamefont {S.~R.}\
  \bibnamefont {Forrest}},\ }\href@noop {} {\bibfield  {journal} {\bibinfo
  {journal} {Phys. Rev. B}\ }\textbf {\bibinfo {volume} {83}},\ \bibinfo
  {pages} {195326} (\bibinfo {year} {2011})}\BibitemShut {NoStop}%
\bibitem [{\citenamefont {Scharber}\ and\ \citenamefont
  {Sariciftci}(2013)}]{Schaber/Sariciftci:2013}%
  \BibitemOpen
  \bibfield  {author} {\bibinfo {author} {\bibfnamefont {M.~C.}\ \bibnamefont
  {Scharber}}\ and\ \bibinfo {author} {\bibfnamefont {N.~S.}\ \bibnamefont
  {Sariciftci}},\ }\href@noop {} {\bibfield  {journal} {\bibinfo  {journal}
  {Prog. Polym. Sci.}\ }\textbf {\bibinfo {volume} {38}},\ \bibinfo {pages}
  {1929} (\bibinfo {year} {2013})}\BibitemShut {NoStop}%
\bibitem [{\citenamefont {Green}(2012)}]{Green:2012}%
  \BibitemOpen
  \bibfield  {author} {\bibinfo {author} {\bibfnamefont {M.~A.}\ \bibnamefont
  {Green}},\ }\href@noop {} {\bibfield  {journal} {\bibinfo  {journal} {Nano
  Letters}\ }\textbf {\bibinfo {volume} {12}},\ \bibinfo {pages} {5985}
  (\bibinfo {year} {2012})}\BibitemShut {NoStop}%
\bibitem [{\citenamefont {Sylvester-Hvid}\ \emph {et~al.}(2004)\citenamefont
  {Sylvester-Hvid}, \citenamefont {Rettrup},\ and\ \citenamefont
  {Ratner}}]{Sylvester-Hivid/etal:2004}%
  \BibitemOpen
  \bibfield  {author} {\bibinfo {author} {\bibfnamefont {K.~O.}\ \bibnamefont
  {Sylvester-Hvid}}, \bibinfo {author} {\bibfnamefont {S.}~\bibnamefont
  {Rettrup}}, \ and\ \bibinfo {author} {\bibfnamefont {M.~A.}\ \bibnamefont
  {Ratner}},\ }\href@noop {} {\bibfield  {journal} {\bibinfo  {journal} {J.
  Chem. B,}\ }\textbf {\bibinfo {volume} {108}},\ \bibinfo {pages} {4296}
  (\bibinfo {year} {2004})}\BibitemShut {NoStop}%
\bibitem [{\citenamefont {Rau}(2007)}]{Rau:2007}%
  \BibitemOpen
  \bibfield  {author} {\bibinfo {author} {\bibfnamefont {U.}~\bibnamefont
  {Rau}},\ }\href@noop {} {\bibfield  {journal} {\bibinfo  {journal} {Phys.
  Rev. B}\ }\textbf {\bibinfo {volume} {76}},\ \bibinfo {pages} {085303}
  (\bibinfo {year} {2007})}\BibitemShut {NoStop}%
\bibitem [{\citenamefont {Kirchartz}\ and\ \citenamefont
  {Rau}(2008)}]{Kirchartz/Rau:2008}%
  \BibitemOpen
  \bibfield  {author} {\bibinfo {author} {\bibfnamefont {T.}~\bibnamefont
  {Kirchartz}}\ and\ \bibinfo {author} {\bibfnamefont {U.}~\bibnamefont
  {Rau}},\ }\href@noop {} {\bibfield  {journal} {\bibinfo  {journal} {physica
  status solidi (a)}\ }\textbf {\bibinfo {volume} {205}},\ \bibinfo {pages}
  {2737} (\bibinfo {year} {2008})}\BibitemShut {NoStop}%
\bibitem [{\citenamefont {Kirchartz}\ \emph {et~al.}(2009)\citenamefont
  {Kirchartz}, \citenamefont {Taretto},\ and\ \citenamefont
  {Rau}}]{Kirchartz/etal:2009a}%
  \BibitemOpen
  \bibfield  {author} {\bibinfo {author} {\bibfnamefont {T.}~\bibnamefont
  {Kirchartz}}, \bibinfo {author} {\bibfnamefont {K.}~\bibnamefont {Taretto}},
  \ and\ \bibinfo {author} {\bibfnamefont {U.}~\bibnamefont {Rau}},\
  }\href@noop {} {\bibfield  {journal} {\bibinfo  {journal} {J. Phys. Chem. C}\
  }\textbf {\bibinfo {volume} {113}},\ \bibinfo {pages} {17958} (\bibinfo
  {year} {2009})}\BibitemShut {NoStop}%
\bibitem [{\citenamefont {Vandewal}\ \emph {et~al.}(2009)\citenamefont
  {Vandewal}, \citenamefont {Tvingstedt}, \citenamefont {Gadisa}, \citenamefont
  {Inganas},\ and\ \citenamefont {Manca}}]{Vandewal/etal:2009}%
  \BibitemOpen
  \bibfield  {author} {\bibinfo {author} {\bibfnamefont {K.}~\bibnamefont
  {Vandewal}}, \bibinfo {author} {\bibfnamefont {K.}~\bibnamefont
  {Tvingstedt}}, \bibinfo {author} {\bibfnamefont {A.}~\bibnamefont {Gadisa}},
  \bibinfo {author} {\bibfnamefont {O.}~\bibnamefont {Inganas}}, \ and\
  \bibinfo {author} {\bibfnamefont {J.~V.}\ \bibnamefont {Manca}},\ }\href@noop
  {} {\bibfield  {journal} {\bibinfo  {journal} {Nat. Mater.}\ }\textbf
  {\bibinfo {volume} {8}},\ \bibinfo {pages} {904} (\bibinfo {year}
  {2009})}\BibitemShut {NoStop}%
\bibitem [{\citenamefont {Miyadera}\ \emph {et~al.}(2014)\citenamefont
  {Miyadera}, \citenamefont {Wang}, \citenamefont {Yamanari}, \citenamefont
  {Matsubara},\ and\ \citenamefont {Yoshida}}]{Miyadera/etal:2014}%
  \BibitemOpen
  \bibfield  {author} {\bibinfo {author} {\bibfnamefont {T.}~\bibnamefont
  {Miyadera}}, \bibinfo {author} {\bibfnamefont {Z.}~\bibnamefont {Wang}},
  \bibinfo {author} {\bibfnamefont {T.}~\bibnamefont {Yamanari}}, \bibinfo
  {author} {\bibfnamefont {K.}~\bibnamefont {Matsubara}}, \ and\ \bibinfo
  {author} {\bibfnamefont {Y.}~\bibnamefont {Yoshida}},\ }\href@noop {}
  {\bibfield  {journal} {\bibinfo  {journal} {Jpn. J. Appl. Phys.}\ }\textbf
  {\bibinfo {volume} {53}},\ \bibinfo {pages} {01AB12} (\bibinfo {year}
  {2014})}\BibitemShut {NoStop}%
\bibitem [{\citenamefont {Nelson}\ \emph {et~al.}(2004)\citenamefont {Nelson},
  \citenamefont {Kirkpatrick},\ and\ \citenamefont
  {Ravirajan}}]{Nelson/etal:2004}%
  \BibitemOpen
  \bibfield  {author} {\bibinfo {author} {\bibfnamefont {J.}~\bibnamefont
  {Nelson}}, \bibinfo {author} {\bibfnamefont {J.}~\bibnamefont {Kirkpatrick}},
  \ and\ \bibinfo {author} {\bibfnamefont {P.}~\bibnamefont {Ravirajan}},\
  }\href@noop {} {\bibfield  {journal} {\bibinfo  {journal} {Phys. Rev. B}\
  }\textbf {\bibinfo {volume} {69}},\ \bibinfo {pages} {035337} (\bibinfo
  {year} {2004})}\BibitemShut {NoStop}%
\bibitem [{\citenamefont {Markvart}(2008)}]{Markvart:2008}%
  \BibitemOpen
  \bibfield  {author} {\bibinfo {author} {\bibfnamefont {T.}~\bibnamefont
  {Markvart}},\ }\href@noop {} {\bibfield  {journal} {\bibinfo  {journal}
  {Phys. Stat. Sol. (a)}\ }\textbf {\bibinfo {volume} {205}},\ \bibinfo {pages}
  {2752} (\bibinfo {year} {2008})}\BibitemShut {NoStop}%
\bibitem [{\citenamefont {Rutten}\ \emph {et~al.}(2009)\citenamefont {Rutten},
  \citenamefont {Esposito},\ and\ \citenamefont {Cleuren}}]{Rutten/etal:2009}%
  \BibitemOpen
  \bibfield  {author} {\bibinfo {author} {\bibfnamefont {B.}~\bibnamefont
  {Rutten}}, \bibinfo {author} {\bibfnamefont {M.}~\bibnamefont {Esposito}}, \
  and\ \bibinfo {author} {\bibfnamefont {B.}~\bibnamefont {Cleuren}},\
  }\href@noop {} {\bibfield  {journal} {\bibinfo  {journal} {Phys. Rev. B}\
  }\textbf {\bibinfo {volume} {80}},\ \bibinfo {pages} {235122} (\bibinfo
  {year} {2009})}\BibitemShut {NoStop}%
\bibitem [{\citenamefont {Einax}\ \emph {et~al.}(2011)\citenamefont {Einax},
  \citenamefont {Dierl},\ and\ \citenamefont {Nitzan}}]{Einax/etal:2011}%
  \BibitemOpen
  \bibfield  {author} {\bibinfo {author} {\bibfnamefont {M.}~\bibnamefont
  {Einax}}, \bibinfo {author} {\bibfnamefont {M.}~\bibnamefont {Dierl}}, \ and\
  \bibinfo {author} {\bibfnamefont {A.}~\bibnamefont {Nitzan}},\ }\href@noop {}
  {\bibfield  {journal} {\bibinfo  {journal} {J. Phys. Chem. C}\ }\textbf
  {\bibinfo {volume} {115}},\ \bibinfo {pages} {21396} (\bibinfo {year}
  {2011})}\BibitemShut {NoStop}%
\bibitem [{\citenamefont {Einax}\ \emph {et~al.}(2013)\citenamefont {Einax},
  \citenamefont {Dierl}, \citenamefont {Schiff},\ and\ \citenamefont
  {Nitzan}}]{Einax/etal:2013}%
  \BibitemOpen
  \bibfield  {author} {\bibinfo {author} {\bibfnamefont {M.}~\bibnamefont
  {Einax}}, \bibinfo {author} {\bibfnamefont {M.}~\bibnamefont {Dierl}},
  \bibinfo {author} {\bibfnamefont {P.~R.}\ \bibnamefont {Schiff}}, \ and\
  \bibinfo {author} {\bibfnamefont {A.}~\bibnamefont {Nitzan}},\ }\href@noop {}
  {\bibfield  {journal} {\bibinfo  {journal} {Europhys. Lett.}\ }\textbf
  {\bibinfo {volume} {104}},\ \bibinfo {pages} {40002} (\bibinfo {year}
  {2013})}\BibitemShut {NoStop}%
\bibitem [{\citenamefont {Wang}\ and\ \citenamefont {Wu}(2012)}]{Wang/Wu:2012}%
  \BibitemOpen
  \bibfield  {author} {\bibinfo {author} {\bibfnamefont {H.}~\bibnamefont
  {Wang}}\ and\ \bibinfo {author} {\bibfnamefont {G.}~\bibnamefont {Wu}},\
  }\href@noop {} {\bibfield  {journal} {\bibinfo  {journal} {Phys. Lett. A}\
  }\textbf {\bibinfo {volume} {376}},\ \bibinfo {pages} {2209 } (\bibinfo
  {year} {2012})}\BibitemShut {NoStop}%
\bibitem [{\citenamefont {Scully}(2010)}]{Scully:2010}%
  \BibitemOpen
  \bibfield  {author} {\bibinfo {author} {\bibfnamefont {M.~O.}\ \bibnamefont
  {Scully}},\ }\href@noop {} {\bibfield  {journal} {\bibinfo  {journal} {Phys.
  Rev. Lett.}\ }\textbf {\bibinfo {volume} {104}},\ \bibinfo {pages} {207701}
  (\bibinfo {year} {2010})}\BibitemShut {NoStop}%
\bibitem [{\citenamefont {Kirk}(2011)}]{Kirk:2011}%
  \BibitemOpen
  \bibfield  {author} {\bibinfo {author} {\bibfnamefont {A.~P.}\ \bibnamefont
  {Kirk}},\ }\href@noop {} {\bibfield  {journal} {\bibinfo  {journal} {Phys.
  Rev. Lett.}\ }\textbf {\bibinfo {volume} {106}},\ \bibinfo {pages} {048703}
  (\bibinfo {year} {2011})}\BibitemShut {NoStop}%
\bibitem [{\citenamefont {Scully}(2011)}]{Scully:2011}%
  \BibitemOpen
  \bibfield  {author} {\bibinfo {author} {\bibfnamefont {M.~O.}\ \bibnamefont
  {Scully}},\ }\href@noop {} {\bibfield  {journal} {\bibinfo  {journal} {Phys.
  Rev. Lett.}\ }\textbf {\bibinfo {volume} {106}},\ \bibinfo {pages} {049801}
  (\bibinfo {year} {2011})}\BibitemShut {NoStop}%
\bibitem [{\citenamefont {Goswami}\ and\ \citenamefont
  {Harbola}(2013)}]{Goswami/Harbola:2013}%
  \BibitemOpen
  \bibfield  {author} {\bibinfo {author} {\bibfnamefont {H.~P.}\ \bibnamefont
  {Goswami}}\ and\ \bibinfo {author} {\bibfnamefont {U.}~\bibnamefont
  {Harbola}},\ }\href@noop {} {\bibfield  {journal} {\bibinfo  {journal} {Phys.
  Rev. A}\ }\textbf {\bibinfo {volume} {88}},\ \bibinfo {pages} {013842}
  (\bibinfo {year} {2013})}\BibitemShut {NoStop}%
\bibitem [{\citenamefont {Hill}(1966)}]{Hill:1966}%
  \BibitemOpen
  \bibfield  {author} {\bibinfo {author} {\bibfnamefont {T.~L.}\ \bibnamefont
  {Hill}},\ }\href@noop {} {\bibfield  {journal} {\bibinfo  {journal} {Journal
  of Theoretical Biology}\ }\textbf {\bibinfo {volume} {10}},\ \bibinfo {pages}
  {442 } (\bibinfo {year} {1966})}\BibitemShut {NoStop}%
\bibitem [{\citenamefont {Schnakenberg}(1976)}]{Schnakenberg:1976}%
  \BibitemOpen
  \bibfield  {author} {\bibinfo {author} {\bibfnamefont {J.}~\bibnamefont
  {Schnakenberg}},\ }\href@noop {} {\bibfield  {journal} {\bibinfo  {journal}
  {Rev. Mod. Phys.}\ }\textbf {\bibinfo {volume} {48}},\ \bibinfo {pages} {571}
  (\bibinfo {year} {1976})}\BibitemShut {NoStop}%
\bibitem [{\citenamefont {Zia}\ and\ \citenamefont
  {Schmittmann}(2007)}]{Zia/Schmittmann:2007}%
  \BibitemOpen
  \bibfield  {author} {\bibinfo {author} {\bibfnamefont {R.}~\bibnamefont
  {Zia}}\ and\ \bibinfo {author} {\bibfnamefont {B.}~\bibnamefont
  {Schmittmann}},\ }\href@noop {} {\bibfield  {journal} {\bibinfo  {journal}
  {J. Stat. Mech.}\ }\textbf {\bibinfo {volume} {P07012}} (\bibinfo {year}
  {2007})}\BibitemShut {NoStop}%
\bibitem [{\citenamefont {Andrieux}\ and\ \citenamefont
  {Gaspard}(2007)}]{Andrieux/Gaspard:2007}%
  \BibitemOpen
  \bibfield  {author} {\bibinfo {author} {\bibfnamefont {D.}~\bibnamefont
  {Andrieux}}\ and\ \bibinfo {author} {\bibfnamefont {P.}~\bibnamefont
  {Gaspard}},\ }\href@noop {} {\bibfield  {journal} {\bibinfo  {journal} {J.
  Stat. Phys.}\ }\textbf {\bibinfo {volume} {127}},\ \bibinfo {pages} {107}
  (\bibinfo {year} {2007})}\BibitemShut {NoStop}%
\bibitem [{\citenamefont {Gaspard}(2010)}]{Gaspard:2010}%
  \BibitemOpen
  \bibfield  {author} {\bibinfo {author} {\bibfnamefont {P.}~\bibnamefont
  {Gaspard}}\ }(\bibinfo  {publisher} {Wiley-VCH},\ \bibinfo {address}
  {Weinheim},\ \bibinfo {year} {2010})\ Chap.\ \bibinfo {chapter} {Nonlinear
  Dynamics of Nanosystems}, p.~\bibinfo {pages} {1}\BibitemShut {NoStop}%
\bibitem [{\citenamefont {Altaner}\ \emph {et~al.}(2012)\citenamefont
  {Altaner}, \citenamefont {Grosskinsky}, \citenamefont {Herminghaus},
  \citenamefont {Katth\"an}, \citenamefont {Timme},\ and\ \citenamefont
  {Vollmer}}]{Altaner/etal:2012}%
  \BibitemOpen
  \bibfield  {author} {\bibinfo {author} {\bibfnamefont {B.}~\bibnamefont
  {Altaner}}, \bibinfo {author} {\bibfnamefont {S.}~\bibnamefont
  {Grosskinsky}}, \bibinfo {author} {\bibfnamefont {S.}~\bibnamefont
  {Herminghaus}}, \bibinfo {author} {\bibfnamefont {L.}~\bibnamefont
  {Katth\"an}}, \bibinfo {author} {\bibfnamefont {M.}~\bibnamefont {Timme}}, \
  and\ \bibinfo {author} {\bibfnamefont {J.}~\bibnamefont {Vollmer}},\
  }\href@noop {} {\bibfield  {journal} {\bibinfo  {journal} {Phys. Rev. E}\
  }\textbf {\bibinfo {volume} {85}},\ \bibinfo {pages} {041133} (\bibinfo
  {year} {2012})}\BibitemShut {NoStop}%
\bibitem [{\citenamefont {Seifert}(2012)}]{Seifert:2012}%
  \BibitemOpen
  \bibfield  {author} {\bibinfo {author} {\bibfnamefont {U.}~\bibnamefont
  {Seifert}},\ }\href@noop {} {\bibfield  {journal} {\bibinfo  {journal} {Rep.
  Prog. Phys.}\ }\textbf {\bibinfo {volume} {75}},\ \bibinfo {pages} {126001}
  (\bibinfo {year} {2012})}\BibitemShut {NoStop}%
\bibitem [{\citenamefont {Kirchhoff}(1847)}]{Kirchhoff:1847}%
  \BibitemOpen
  \bibfield  {author} {\bibinfo {author} {\bibfnamefont {G.}~\bibnamefont
  {Kirchhoff}},\ }\href@noop {} {\bibfield  {journal} {\bibinfo  {journal}
  {Ann. Phys. (Berlin)}\ }\textbf {\bibinfo {volume} {148}},\ \bibinfo {pages}
  {497} (\bibinfo {year} {1847})}\BibitemShut {NoStop}%
\bibitem [{\citenamefont {Andrieux}\ and\ \citenamefont
  {Gaspard}(2004)}]{Andrieux/Gaspard:2004}%
  \BibitemOpen
  \bibfield  {author} {\bibinfo {author} {\bibfnamefont {D.}~\bibnamefont
  {Andrieux}}\ and\ \bibinfo {author} {\bibfnamefont {P.}~\bibnamefont
  {Gaspard}},\ }\href@noop {} {\bibfield  {journal} {\bibinfo  {journal} {J.
  Chem. Phys.}\ }\textbf {\bibinfo {volume} {121}},\ \bibinfo {pages} {6167}
  (\bibinfo {year} {2004})}\BibitemShut {NoStop}%
\bibitem [{\citenamefont {Gerritsma}\ and\ \citenamefont
  {Gaspard}(2010)}]{Gerritsma/Gaspard:2010}%
  \BibitemOpen
  \bibfield  {author} {\bibinfo {author} {\bibfnamefont {E.}~\bibnamefont
  {Gerritsma}}\ and\ \bibinfo {author} {\bibfnamefont {P.}~\bibnamefont
  {Gaspard}},\ }\href@noop {} {\bibfield  {journal} {\bibinfo  {journal}
  {Biophysical Reviews and Letters}\ }\textbf {\bibinfo {volume} {05}},\
  \bibinfo {pages} {163} (\bibinfo {year} {2010})}\BibitemShut {NoStop}%
\bibitem [{\citenamefont {Einax}\ \emph
  {et~al.}(2010{\natexlab{a}})\citenamefont {Einax}, \citenamefont {Solomon},
  \citenamefont {Dieterich},\ and\ \citenamefont {Nitzan}}]{Einax/etal:2010a}%
  \BibitemOpen
  \bibfield  {author} {\bibinfo {author} {\bibfnamefont {M.}~\bibnamefont
  {Einax}}, \bibinfo {author} {\bibfnamefont {G.~C.}\ \bibnamefont {Solomon}},
  \bibinfo {author} {\bibfnamefont {W.}~\bibnamefont {Dieterich}}, \ and\
  \bibinfo {author} {\bibfnamefont {A.}~\bibnamefont {Nitzan}},\ }\href@noop {}
  {\bibfield  {journal} {\bibinfo  {journal} {J. Chem. Phys.}\ }\textbf
  {\bibinfo {volume} {133}},\ \bibinfo {pages} {054102} (\bibinfo {year}
  {2010}{\natexlab{a}})}\BibitemShut {NoStop}%
\bibitem [{\citenamefont {Einax}\ \emph
  {et~al.}(2010{\natexlab{b}})\citenamefont {Einax}, \citenamefont {K\"orner},
  \citenamefont {Maass},\ and\ \citenamefont {Nitzan}}]{Einax/etal:2010b}%
  \BibitemOpen
  \bibfield  {author} {\bibinfo {author} {\bibfnamefont {M.}~\bibnamefont
  {Einax}}, \bibinfo {author} {\bibfnamefont {M.}~\bibnamefont {K\"orner}},
  \bibinfo {author} {\bibfnamefont {P.}~\bibnamefont {Maass}}, \ and\ \bibinfo
  {author} {\bibfnamefont {A.}~\bibnamefont {Nitzan}},\ }\href@noop {}
  {\bibfield  {journal} {\bibinfo  {journal} {Phys. Chem. Chem. Phys.}\
  }\textbf {\bibinfo {volume} {12}},\ \bibinfo {pages} {645} (\bibinfo {year}
  {2010}{\natexlab{b}})}\BibitemShut {NoStop}%
\bibitem [{\citenamefont {Dierl}\ \emph {et~al.}(2011)\citenamefont {Dierl},
  \citenamefont {Maass},\ and\ \citenamefont {Einax}}]{Dierl/etal:2011}%
  \BibitemOpen
  \bibfield  {author} {\bibinfo {author} {\bibfnamefont {M.}~\bibnamefont
  {Dierl}}, \bibinfo {author} {\bibfnamefont {P.}~\bibnamefont {Maass}}, \ and\
  \bibinfo {author} {\bibfnamefont {M.}~\bibnamefont {Einax}},\ }\href@noop {}
  {\bibfield  {journal} {\bibinfo  {journal} {Europhys. Lett.}\ }\textbf
  {\bibinfo {volume} {93}},\ \bibinfo {pages} {50003} (\bibinfo {year}
  {2011})}\BibitemShut {NoStop}%
\bibitem [{\citenamefont {Dierl}\ \emph {et~al.}(2012)\citenamefont {Dierl},
  \citenamefont {Maass},\ and\ \citenamefont {Einax}}]{Dierl/etal:2012}%
  \BibitemOpen
  \bibfield  {author} {\bibinfo {author} {\bibfnamefont {M.}~\bibnamefont
  {Dierl}}, \bibinfo {author} {\bibfnamefont {P.}~\bibnamefont {Maass}}, \ and\
  \bibinfo {author} {\bibfnamefont {M.}~\bibnamefont {Einax}},\ }\href@noop {}
  {\bibfield  {journal} {\bibinfo  {journal} {Phys. Rev. Lett.}\ }\textbf
  {\bibinfo {volume} {108}},\ \bibinfo {pages} {060603} (\bibinfo {year}
  {2012})}\BibitemShut {NoStop}%
\bibitem [{\citenamefont {Burlakov}\ \emph {et~al.}(2005)\citenamefont
  {Burlakov}, \citenamefont {Kawata}, \citenamefont {Assender}, \citenamefont
  {Briggs}, \citenamefont {Rusecks},\ and\ \citenamefont
  {Samuel}}]{Burlakov/etal:2005}%
  \BibitemOpen
  \bibfield  {author} {\bibinfo {author} {\bibfnamefont {V.~M.}\ \bibnamefont
  {Burlakov}}, \bibinfo {author} {\bibfnamefont {K.}~\bibnamefont {Kawata}},
  \bibinfo {author} {\bibfnamefont {H.~E.}\ \bibnamefont {Assender}}, \bibinfo
  {author} {\bibfnamefont {G.~A.~D.}\ \bibnamefont {Briggs}}, \bibinfo {author}
  {\bibfnamefont {A.}~\bibnamefont {Rusecks}}, \ and\ \bibinfo {author}
  {\bibfnamefont {I.~D.~W.}\ \bibnamefont {Samuel}},\ }\href@noop {} {\bibfield
   {journal} {\bibinfo  {journal} {Phys. Rev. B}\ }\textbf {\bibinfo {volume}
  {72}},\ \bibinfo {pages} {075206} (\bibinfo {year} {2005})}\BibitemShut
  {NoStop}%
\bibitem [{\citenamefont {R{\"u}hle}\ \emph {et~al.}(2011)\citenamefont
  {R{\"u}hle}, \citenamefont {Lukyanov}, \citenamefont {May}, \citenamefont
  {Schrader}, \citenamefont {Vehoff}, \citenamefont {Kirkpatrick},
  \citenamefont {Baumeier},\ and\ \citenamefont
  {Andrienko}}]{Ruehle/etal:2011}%
  \BibitemOpen
  \bibfield  {author} {\bibinfo {author} {\bibfnamefont {V.}~\bibnamefont
  {R{\"u}hle}}, \bibinfo {author} {\bibfnamefont {A.}~\bibnamefont {Lukyanov}},
  \bibinfo {author} {\bibfnamefont {F.}~\bibnamefont {May}}, \bibinfo {author}
  {\bibfnamefont {M.}~\bibnamefont {Schrader}}, \bibinfo {author}
  {\bibfnamefont {T.}~\bibnamefont {Vehoff}}, \bibinfo {author} {\bibfnamefont
  {J.}~\bibnamefont {Kirkpatrick}}, \bibinfo {author} {\bibfnamefont
  {B.}~\bibnamefont {Baumeier}}, \ and\ \bibinfo {author} {\bibfnamefont
  {D.}~\bibnamefont {Andrienko}},\ }\href@noop {} {\bibfield  {journal}
  {\bibinfo  {journal} {Journal of Chemical Theory and Computation}\ }\textbf
  {\bibinfo {volume} {7}},\ \bibinfo {pages} {3335} (\bibinfo {year}
  {2011})}\BibitemShut {NoStop}%
\bibitem [{\citenamefont {Baruch}(1985)}]{Baruch:1985}%
  \BibitemOpen
  \bibfield  {author} {\bibinfo {author} {\bibfnamefont {P.}~\bibnamefont
  {Baruch}},\ }\href@noop {} {\bibfield  {journal} {\bibinfo  {journal} {J.
  Appl. Phys.}\ }\textbf {\bibinfo {volume} {57}},\ \bibinfo {pages} {1347}
  (\bibinfo {year} {1985})}\BibitemShut {NoStop}%
\bibitem [{not({\natexlab{a}})}]{note:EN_1}%
  \BibitemOpen
  \href@noop {} {} \ \bibinfo {note} {obviously, as long as
  the pumping is represented by kinetic rates, an effective temperature can be
  defined. It is often referred to as ``sun temperature'', however it is
  usually a more complex quantity as some of the processes that cause
  transition between states $1$ and $2$, for example non-radiative relaxation
  (i.~e. recombination) are associated with the ambient
  environment.}\BibitemShut {Stop}%
\bibitem [{not({\natexlab{b}})}]{note:EN_2}%
  \BibitemOpen
  \href@noop {} {} \ \bibinfo {note} {because level $A_1$ is
  restricted to be always occupied, the single electron energy
  $\varepsilon_{\rm\scriptscriptstyle A2}$ is taken to include the coulombic
  repulsion between two electrons on the acceptor. This notation is different
  from that of Refs.~\onlinecite{Einax/etal:2011,Einax/etal:2013}, where we
  have referred to this repulsive interaction explicitly.}\BibitemShut {Stop}%
\bibitem [{not({\natexlab{c}})}]{note:EN_3}%
  \BibitemOpen
  \href@noop {} {} \ \bibinfo {note} {in this case the maximum
  achievable efficiency value for the chosen parameter set depends only on the
  detailed balance ratio and is not affected by the specific form of the
  transition rates for the electron hopping between the reservoirs and the
  molecule sites as well as between the donor and acceptor
  molecules.}\BibitemShut {Stop}%
\bibitem [{\citenamefont {Einax}\ and\ \citenamefont
  {Nitzan}()}]{Einax/Nitzan:2014}%
  \BibitemOpen
  \bibfield  {author} {\bibinfo {author} {\bibfnamefont {M.}~\bibnamefont
  {Einax}}\ and\ \bibinfo {author} {\bibfnamefont {A.}~\bibnamefont {Nitzan}},\
  }\href@noop {} {}\bibinfo {note} {to be published}\BibitemShut {NoStop}%
\end{thebibliography}%

\end{document}